\overfullrule=0pt
\input harvmac

\lref\BerkovitsBT{
  N.~Berkovits,
  ``Pure spinor formalism as an N=2 topological string,''
JHEP {\bf 0510}, 089 (2005).
[hep-th/0509120].
}

\lref\BerkovitsXU{
  N.~Berkovits,
  ``Quantum consistency of the superstring in AdS(5) x S**5 background,''
JHEP {\bf 0503}, 041 (2005).
[hep-th/0411170].
}

\lref\BerkovitsZK{
  N.~Berkovits,
  ``ICTP lectures on covariant quantization of the superstring,''
[hep-th/0209059].
}

\lref\sym{
  E.~Witten,
  ``Perturbative gauge theory as a string theory in twistor space,''
Commun.\ Math.\ Phys.\  {\bf 252}, 189 (2004).
[hep-th/0312171]\semi
N.~Berkovits,
  ``An Alternative string theory in twistor space for N=4 superYang-Mills,''
Phys.\ Rev.\ Lett.\  {\bf 93}, 011601 (2004).
[hep-th/0402045].}

\lref\tentw{
  L.~Mason and D.~Skinner,
  ``Ambitwistor strings and the scattering equations,''
JHEP {\bf 1407}, 048 (2014).
[arXiv:1311.2564 [hep-th]]\semi
  N.~Berkovits,
  ``Infinite Tension Limit of the Pure Spinor Superstring,''
JHEP {\bf 1403}, 017 (2014).
[arXiv:1311.4156 [hep-th]].
}

\lref\BerkovitsFE{
  N.~Berkovits,
  ``Super Poincare covariant quantization of the superstring,''
JHEP {\bf 0004}, 018 (2000).
[hep-th/0001035].
}

\lref\memem{M.~Cederwall, B.~E.~W.~Nilsson and D.~Tsimpis,
  ``Spinorial cohomology and maximally supersymmetric theories,''
JHEP {\bf 0202}, 009 (2002).
[hep-th/0110069].}

\lref\bmem{ N.~Berkovits,
  ``Towards covariant quantization of the supermembrane,''
JHEP {\bf 0209}, 051 (2002).
[hep-th/0201151].}

\lref\cmem{M.~Cederwall,
  ``D=11 supergravity with manifest supersymmetry,''
Mod.\ Phys.\ Lett.\ A {\bf 25}, 3201 (2010).
[arXiv:1001.0112 [hep-th]].
}

\lref\town{
  E.~Bergshoeff, E.~Sezgin and P.~K.~Townsend,
  ``Supermembranes and Eleven-Dimensional Supergravity,''
Phys.\ Lett.\ B {\bf 189}, 75 (1987).
}

\lref\BH{
  N.~Berkovits and P.~S.~Howe,
  ``Ten-dimensional supergravity constraints from the pure spinor formalism for the superstring,''
Nucl.\ Phys.\ B {\bf 635}, 75 (2002).
[hep-th/0112160].
}

\lref\SiegelYD{
  W.~Siegel,
[arXiv:1005.2317 [hep-th]].
}

\def\bar{\overline}

\def\a{{\alpha}}

\def\ah{{\hat \a}}

\def\lh{{\widehat \lambda}}

\def\kh{{\widehat \kappa}}

\def\wh{{\widehat w}}

\def\l{{\lambda}}

\def\b{{\beta}}
\def\bh{{\hat\beta}}

\def\g{{\gamma}}

\def\d{{\delta}}
\def\e{{\epsilon}}
\def\s{{\sigma}}
\def\k{{\kappa}}

\def\N{{\nabla}}
\def\Nh{{\widehat\nabla}}
\def\L{{\Lambda}}

\def\O{{\Omega}}
\def\Oh{{\widehat\Omega}}
\def\wh{{\widehat w}}

\def\half{{1\over 2}}
\def\p{{\partial}}

\def\t{{\theta}}

\def\th{{\widehat\theta}}

\Title{\vbox{\baselineskip12pt
\hbox{}}}
{{\vbox{\centerline{Origin of the Pure Spinor and}
\smallskip
\centerline{ Green-Schwarz Formalisms
 }
}} }
\bigskip\centerline{Nathan Berkovits\foot{e-mail: nberkovi@ift.unesp.br}}
\bigskip
\centerline{\it ICTP South American Institute for Fundamental Research 
}
\centerline{\it Instituto de F\'\i sica Te\'orica, UNESP - Univ. 
Estadual Paulista }
\centerline{\it Rua Dr. Bento T. Ferraz 271, 01140-070, S\~ao Paulo, SP, Brasil}
\smallskip

\vskip .3in

The pure spinor formalism for the superstring was recently obtained by gauge-fixing
a purely bosonic classical action involving a twistor-like constraint $\partial x^m (\gamma_m\lambda)_\alpha =0$ where $\lambda^\alpha$ is a d=10 pure spinor. This twistor-like constraint replaces the usual Virasoro constraint $\partial x^m \partial x_m =0$, and the  Green-Schwarz fermionic spacetime spinor variables $\theta^\alpha$ arise as Faddeev-Popov ghosts for this constraint.

In this paper, the purely bosonic classical action is simplified by replacing the classical d=10 pure spinor $\lambda^\alpha$ with a d=10 projective pure spinor. The pure spinor and Green-Schwarz formalisms for the superparticle and superstring are then obtained as different gauge-fixings of this purely bosonic classical action, and the Green-Schwarz  kappa symmetry is directly related to the pure spinor BRST symmetry. Since a d=10 projective pure spinor parameterizes ${{SO(10)}\over{U(5)}}$, this action can be interpreted as a standard $\hat c=5$ topological
action where one integrates over the ${{SO(10)}\over{U(5)}}$ choice of complex structure.
Finally, a purely bosonic action for the d=11 supermembrane is proposed which
reduces upon double-dimensional reduction to the purely bosonic action for the d=10 Type IIA superstring.

\vskip .3in

\Date {March 2015}

\newsec{Introduction}

In the conventional Ramond-Neveu-Schwarz (RNS) formalism for the superstring, the
fermionic worldsheet variables $\psi^m$ are d=10 spacetime vectors and d=2 worldsheet
spinors. From the spacetime point of view, a d=10 vector representation for fermionic variables is unusual and implies that spacetime supersymmetry acts in a complicated manner mixing $\psi^m$ with
ghost variables, and that
Ramond-Ramond backgrounds cannot be described in a straightforward manner. However, from
the worldsheet point of view, their d=2 spinor representation is consistent with spin-statistics
and $\psi^m$ transforms in a natural manner under worldsheet supersymmetry
in the classical RNS
worldsheet action. 
 
On the other hand, in the Green-Schwarz (GS) and pure spinor formalisms for the superstring, the
fermionic worldsheet variables $\t^\a$ are d=10 spacetime spinors and d=2 worldsheet
scalars. In this case, their d=10 spacetime spinor representation is natural so spacetime supersymmetry acts covariantly and there is no problem in describing Ramond-Ramond
backgrounds. However, their d=2 worldsheet scalar representation is in conflict with spin-statistics, and $\t^\a$ transforms in an unusual manner under the
fermionic worldsheet symmetry called kappa symmetry in the GS formalism or BRST symmetry
in the pure spinor formalism. 

In this paper, a new interpretation for the GS and pure spinor formalisms will be proposed in which the
fermionic $\t^\a$ variables are not classical variables but arise as Faddeev-Popov ghosts from gauge-fixing a twistor-like action.
The classical action will be constructed only from bosonic variables which include the
usual d=10 spacetime vector variable $x^m$ together with a d=10 projective pure spinor variable
$\l^\a$ that plays the role of a twistor variable. Since a d=10 projective pure spinor parameterizes ${{SO(10)}\over{U(5)}}$, this action can be interpreted as a standard $\hat c=5$ topological
action where one integrates over the ${{SO(10)}\over{U(5)}}$ choice of complex structure.
Different gauge-fixings of this purely bosonic classical action will produce either the pure 
spinor or GS formalisms, and the GS kappa symmetry will be related in
a simple manner to the pure spinor BRST symmetry. 

For the superparticle, the spacetime momentum $P_m$ will
satisfy the twistor-like constraint $P_m (\g^m \l)_\a =0$ which replaces the mass-shell constraint $P_m P^m=0$, and the fermionic variable $\t^\a$ is the Faddeev-Popov ghost for this constraint which is a worldline scalar and spacetime spinor. For the Type II superstring, the twistor-like
constraints will be 
\eqn\twc{(P_m + \partial_\s x_m)(\g^m\l)_\a =0 \quad{\rm and} \quad
(P_m - \partial_\s x_m)(\g^m\lh)_\ah =0}
which replace the Virasoro
constraints $(P_m + \partial_\s x_m)(P^m + \partial_\s x^m)=0$ and $(P_m - \partial_\s x_m)(P^m - \partial_\s x^m)=0$, and the fermionic Faddeev-Popov ghosts $\t^\a$ and $\th^\ah$ for these constraints are worldsheet scalars and d=10 spacetime spinors. Finally, for the supermembrane,  the twistor-like
constraint will be $P_M (\g^M\l)_B + \partial_{\s_1} x_M \partial_{\s_2} x_N (\g^{MN}\l)_B =0$ 
which replaces the reparameterization 
constraints $P_M P^M + det (\p_{\s_j} x^M \p_{\s_k} x_M)=0$ and $P_M  \partial_{\s_j} x^M=0$, and the fermionic Faddeev-Popov ghost $\t^B$ is a worldvolume scalar and d=11 spacetime spinor. 

After gauge-fixing the bosonic twistor-like action, spacetime supersymmetry mixes matter
and ghost variables in a manner reminiscent of the mixing of matter and ghost variables under spacetime supersymmetry in the RNS formalism. However, unlike the RNS formalism where spacetime supersymmetry acts in a complicated manner, spacetime supersymmetry now acts linearly on the matter and ghost variables and it is straightforward to describe Ramond-Ramond backgrounds.

In previous papers \ref\emergent{N.~Berkovits,
  ``Pure spinors, twistors, and emergent supersymmetry,''
JHEP {\bf 1212}, 006 (2012).
[arXiv:1105.1147 [hep-th]].} and  \ref\recent{ N.~Berkovits,
  ``Twistor Origin of the Superstring,''
[arXiv:1409.2510 [hep-th]].},
a similar purely bosonic classical action was used to derive the pure spinor
formalism for the superstring by gauge-fixing a twistor-like
constraint. However, unlike these previous papers in which the classical variable $\l^\a$ was a d=10 pure spinor, the classical variable $\l^\a$ in this paper will be a projective d=10 pure spinor. The scale part of $\l^\a$ will be a ghost variable coming from gauge-fixing, which will
imply that the scale-invariant fermionic variables $\t^\a$ carry zero ghost number as expected.
Furthermore, the gauge-fixing procedure in this paper will be more straightforward than in these previous two papers.

More explicitly, under the scale transformation $\l^\a \to \Omega\l^\a$, the Lagrange multiplier
$L_\a$ for the twistor-like constraint transforms as $L_\a \to \Omega^{-1} L_\a$ so
the resulting fermionic ghost $\t^\a$ of ghost-number $+1$ will also transform as $\t^\a \to \Omega^{-1}\t^\a$. 
The fermionic ghost $\t^\a$ will have a gauge symmetry $\d\t^\a = \phi\l^\a$ where $\phi$
is a bosonic ghost-for-ghost which transforms as $\phi \to\Omega^{-2}\phi$.
The scale invariance will be fixed by setting $\phi =1$, which implies that $e^{\half \phi} \l^\a$ and $e^{-\half\phi} \t^\a$ are scale-invariant variables. Since $\phi$ has
ghost-number $+2$, these scale-invariant versions of the $\l^\a$ and $\t^\a$ variables
carry ghost-number $+1$ and 0 respectively.
 
The connection of twistors and the Green-Schwarz and pure spinor formalisms has a long history starting
with \ref\witten{E.~Witten, ``Twistor - Like Transform in Ten-Dimensions,''
Nucl.\ Phys.\ B {\bf 266}, 245 (1986).} which related GS kappa symmetry with
integrability along light-line lines and \ref\howe{ P.~S.~Howe,
  ``Pure spinors lines in superspace and ten-dimensional supersymmetric theories,''
Phys.\ Lett.\ B {\bf 258}, 141 (1991), [Addendum-ibid.\ B {\bf 259}, 511 (1991)].
} which related pure spinor BRST invariance with integrability along pure spinor lines. Actions involving twistor-like variables have
been constructed for the GS superparticle
\ref\gssuperp{A. Bengtsson, I. Bengtsson, M. Cederwall and N. Linden,
``Particles, superparticles and twistors,'' Phys. Rev. D36, 1766 (1987)\semi
E. Sokatchev, ``Harmonic superparticle,'' Class. Quant. Grav. 4, 237 (1987)\semi
D. Sorokin, V. Tkach and D. Volkov, ``Superparticles,
twistors and Siegel symmetry,'' Mod. Phys. Lett. A4, 901 (1989)\semi
N. Berkovits, ``A supertwistor description of the massless superparticle
in ten-dimensional superspace,'' Nucl. Phys. B350, 193 (1991).} and GS
superstring \ref\gssupers{M.~Tonin,
  ``World sheet supersymmetric formulations of Green-Schwarz superstrings,''
Phys.\ Lett.\ B {\bf 266}, 312 (1991)\semi
  N.~Berkovits,
  ``The Heterotic Green-Schwarz superstring on an N=(2,0) superworldsheet,''
Nucl.\ Phys.\ B {\bf 379}, 96 (1992).
[hep-th/9201004]\semi F.~Delduc, A.~Galperin, P.~S.~Howe and E.~Sokatchev,
  ``A Twistor formulation of the heterotic D = 10 superstring with manifest (8,0) world sheet supersymmetry,''
Phys.\ Rev.\ D {\bf 47}, 578 (1993).
[hep-th/9207050].
}, and more recently,
twistor-string actions have been constructed for N=4 d=4 super-Yang-Mills \sym\
and d=10 super-Yang-Mills and supergravity \tentw. However, all of these twistor-string
actions contain classical fermonic variables and their relation to the action of this paper
is unclear.

It would be very interesting to generalize the procedure of this paper to curved target-space
backgrounds. For NS-NS backgrounds, the obvious guess is to replace the twistor-like
constraints of \twc\ with
\eqn\twcc{(g^{mn}(x) P_n + \partial_\s x^m)E^c_m(x) (\g_c\l)_\a =0 \quad{\rm and} \quad
(g^{mn}(x) P_n - \partial_\s x^m)E^c_m(x) (\g_c\lh)_\ah =0,}
where $E^c_m(x)$ is the vierbein satisfying $g^{mn} E_m^c E_n^d = \eta^{cd}$ and $c,d=0$ to 9 are tangent-space indices.
Constructing a classical twistor-like action for R-R backgrounds is more challenging since the fermionic $\t^\a$ variables only appear after gauge-fixing.

In section 2 of this paper, the bosonic twistor-like action for the d=10 superparticle will be constructed,
and its gauge-fixing to the pure
spinor and GS formalisms will be explained.
In section 3, this bosonic twistor-like action will be generalized to the d=10 superstring,
and its gauge-fixing to the pure
spinor and GS formalisms will be explained.
And finally in section 4, a twistor-like action for the d=11 supermembrane will be constructed which
reduces upon double-dimensional reduction to the twistor-like action for the d=10 Type IIA 
superstring.

\newsec{Twistor-Like Superparticle}

Before discussing the d=10 superstring, it will be useful to first discuss the d=10 superparticle and show how its twistor-like action can be interpreted as an action for a topological particle in which one integrates over
the choices of complex structure.

\subsec{Topological particle}

The classical action for a d=10 topological particle is 
\eqn\toppart{S_c = \int d\tau [P_m \dot x^m - L^a P_a] = \int d\tau [P_a \dot x^a + P_{\bar a} \dot  x^{\bar a} - L^a P_a]}
where $\dot x^m = {{\p}\over{\p\tau}}x^m$, $m=0$ to 9 are d=10 vector indices, $a,\bar a=1$ to 5 labels a complex split of
$x^m \to (x^a, x^{\bar a})$ and $P_m \to (P_a. P_{\bar a})$, and $L^a$ is a Lagrange multiplier imposing the constraint $P_a=0$. For the action to be real, one should choose spacetime signature $(5,5)$, however, one can easily Wick-rotate to any other d=10 signature. For convenience, we will use Euclidean signature $(10,0)$ for the rest of this paper which implies that $x^a$ is the complex conjugate of $x^{\bar a}$.

The action of \toppart\  is invariant under the gauge transformations
\eqn\topgt{\d x^a = \t^a, \quad \d L^a =  \dot \t^a}
for arbitary gauge parameters $\t^a(\tau)$, 
and if one gauge-fixes $L^a=0$ using the standard BRST method, $\t^a$ and its conjugate momentum $p_a$ are interpreted
as fermionic Faddeev-Popov ghosts and antighosts associated to this gauge-fixing. The resulting action
is 
\eqn\topbrst{ S = S_c + \int d\tau Q( p_a L^a) 
 = \int d\tau [ P_m \dot x^m  - p_a \dot \t^a + L^a (M_a - P_a) ]}
where $\chi= p_a L^a$ is the gauge-fixing fermion, $Q = \t^a P_a$ is the BRST operator which generates the transformation of \topgt, and $M_a \equiv Q p_a$ is the Nakanishi-Lautrup field whose auxiliary equation of motion is $M_a = P_a$.
Physical states in the BRST cohomology are easily shown to be anti-holomorphic functions
$V = f(x^{\bar a})$ which satisfy $P_a V =0$.

Although this topological theory is not SO(10) invariant, one can write the action of \toppart\ using ten-dimensional notation as
\eqn\topparttwo{S_c = \int d\tau [P_m \dot x^m - (\xi\g^m L) P_m]}
where $\xi^\a$ for $\a=1$ to 16 is a fixed d=10 pure spinor which chooses an ${{SO(10)}\over{U(5)}}$
complex structure. It will now be shown that if one makes the theory SO(10) invariant
by integrating over the choice of ${{SO(10)}\over{U(5)}}$ complex structure, the theory
is no longer topological and describes the d=10 super-Maxwell theory coming from
quantizing the d=10 superparticle.

\subsec{Worldline action}

To integrate over ${{SO(10)}\over{U(5)}}$ complex structures, one should replace the fixed
$\xi^\a$
in \topparttwo\ with a projective pure spinor variable $\l^\a (\tau)$ satisfying the pure spinor condition
\eqn\projpure{ \l^\a (\tau) \g^m_{\a\b} \l^\b(\tau) =0}
and which is defined up to the local scale transformation
\eqn\scale{  \l^\a(\tau) \sim \Omega(\tau)\l^\a(\tau). }
The conjugate momentum to $\l^\a$ will be called $w_\a$ and is defined
up to the gauge and scale transformations
\eqn\projw{ w_\a \sim w_\a + \Lambda^m (\g^m \l)_\a \quad {\rm and} \quad
w_\a \sim \Omega^{-1} w_\a.} 
In terms of $\l^\a$, the topological constraint $P_a=0$ can now be expressed covariantly as
the twistor-like constraint $P_m (\g^m \l)_\a =0$, and 
the corresponding classical worldline action is
\eqn\partact{S_c = \int d\tau [P_m \dot x^m + w_\a \nabla \l^\a -\half L^\a P_m (\g^m \l)_\a]}
where $\nabla \l^\a = \dot\l^\a + A \l^\a$ and $A$ is a worldline gauge field transforming as
$A \to A - \O^{-1} \dot \O$ under the local scale
transformation 
\eqn\gaugeA{  \l^\a \to \O \l^\a, \quad w_\a \to \O^{-1} w_\a,
\quad L^\a \to \O^{-1} L^\a.} 

In addition to the local scale invariance of \gaugeA, \partact\ is also invariant under the local gauge transformations related to the first-class constraint $P_m (\g^m \l)_\a=0$
\eqn\brstpart{\d L^\a = \nabla \t^\a - a \l^\a, \quad \d x^m =\half \l\g^m \t, \quad \d w_\a =-\half P_m (\g^m \t)_\a}
where $\t^\a(\tau)$ and $a(\tau)$ are arbitrary gauge parameters.\foot{In addition to the symmetries of \brstpart, \partact\ is also invariant under the transformation $\d L^\a = 
a^{mn} (\g_{mn}\l)^\a$ for arbitrary gauge parameter $a^{mn}$. However, this symmetry
does not need to be gauge-fixed if one assumes (as will be explained in the following subsection) that the ghost-for-ghost $\phi$ is non-vanishing. Including this BRST transformation
of $L^\a$ would modify the gauge-fixed action of the following subsection
to include the additonal term
$\int d\tau   p_\a (\g_{mn}\l)^\a a^{mn} $ where $p_\a$ is the antighost. But this additional
term is BRST-trivial since it can be expressed as $\{ Q, \int d\tau [- \half (p \g_{mn}\t)-\half (w\g_{mn}\l)+{1\over 4} \phi^{-1} (\t\g_{mnp}\t) P^p ] \phi^{-1} a^{mn} \}$.} 
Furthermore, one has the local gauge-for-gauge symmetries which transform the gauge parameters as
\eqn\gfg{\d'\t^\a = \phi \l^\a, \quad \d' a = \nabla \phi, \quad \d' w_\a = - \phi {{\d \chi}\over{\d L^\a}}}
where $\phi(\tau)$ is a gauge-for-gauge parameter and $\chi$ is the gauge-fixing fermion. The transformation of $w_\a$ in \gfg\ is necessary
since $\d L^\a$ of \brstpart\ transforms under \gfg\ as $\d'(\d L^\a) =\phi \nabla \l^\a$ which is proportional to the equation of motion for $w_\a$. So if $L^\a$ appears in the gauge-fixing fermion $\chi$, the term $Q\chi$ in the action will transform under \gfg\  as
\eqn\tranchi{\d'(Q\chi) =  {{\d \chi}\over{\d L^\a}}
\d'(\d L^\a) = {{\d \chi}\over{\d L^\a}} \phi \N\l^\a,}
which must be cancelled by a
shift of $w_\a \to w_\a -\phi {{\d \chi}\over{\d L^\a}} $.  

\subsec{Pure spinor superparticle}

In this section, the pure spinor description of the d=10 superparticle will be obtained by
gauge-fixing the twistor-like action of \partact. The first step is to use the gauge symmetries of \brstpart\ and the gauge-for-gauge symmetries of \gfg\ to gauge-fix $L^\a=0$ and $a=0$. Using the BRST method where the BRST operator $Q$ generates the symmetries of \brstpart\ and \gfg, this gauge-fixing is accomplished by adding to the classical action $S_c$ of \partact\ the BRST-trivial
term
\eqn\gfpart{S= S_c +\int d\tau Q (p_\a L^\a + \b a) =S_c + 
\int d\tau [ - p_\a (\nabla \t^\a + a \l^\a) + \b \nabla\phi +M_\a L^\a + N a ]}
where $\t^\a$ and $a$ are fermionic ghosts, $\phi$ is a bosonic ghost-for-ghost, 
$p_\a$ is a fermionic antighost, $\b$ is a bosonic antighost-for-ghost, and $M_\a$ and $N$
are bosonic and fermionic Nakanishi-Lautrup fields defined by $M_\a\equiv Q p_\a$ and
$N\equiv Q\b $. Note that $(\t^\a, p_\a)$ and $(\phi, \b)$ scale under \gaugeA\ as
\eqn\scalenew{\t^\a \to \O^{-1}\t^\a, \quad p_\a \to \O p_\a, \quad \phi \to \O^{-2}\phi, \quad
\b\to \O^2 \b.}
After solving the auxiliary equations of motion of $M_\a$ and $N$, the gauge-fixed action of \gfpart\ is
\eqn\gfapart{S = \int d\tau [ P_m \dot x^m + w_\a \dot \l^\a - p_\a \dot \t^\a + \b \dot\phi
+ A (w_\a \l^\a + p_\a \t^\a - 2\b\phi) ].}

The next step is to gauge-fix the local scale symmetry of \gaugeA. Although the naive choice
would be to gauge-fix one of the 16 components of $\l^\a$ to be equal to 1, this gauge-fixing
would break manifest Lorentz invariance and would only be possible if one component of 
$\l^\a$ were required to be non-vanishing on the entire worldline. Instead of this gauge-fixing of $\l^\a$ which would break Lorentz invariance, the local scale symmetry will be gauge-fixed by scaling
the ghost-for-ghost variable $\phi$ to be equal to 1. This gauge choice manifestly preserves Lorentz invariance, but requires that $\phi$ is non-vanishing on the entire worldline.

Since $\phi$ carries scale weight $-2$ and ghost number $+2$, this gauge-fixing of scale
symmetry means that the ghost number of any gauge-fixed operator is shifted by its scale
weight. In other words, if an operator ${\cal O}_{g,s}$ carries ghost number $g$ and
scale weight $s$, the gauge-fixed version of ${\cal O}_{g,s}$ is the scale-invariant operator $\phi^{s\over 2} {\cal O}_{g,s}$ which carries
ghost-number $g+s$. So after gauge-fixing the scale symmetry, the ghost number of the variables
$(\l^\a, w_\a)$ is shifted from $(0,0)$ to $(+1, -1)$  and the ghost number of the
variables $(\t^\a, p_\a)$ is shifted from $(+1, -1)$ to $(0,0)$. These are the appropriate ghost numbers in the pure spinor formalism for defining physical states.
Furthermore, after gauge-fixing $\phi=1$, the variables $A$ and $\b$ of \gfapart\ satisfy the
auxiliary equations of motion $A=0$ and $\b= \half(w_\a \l^\a + p_\a \t^\a)$, and the action
of \gfapart\ reduces to the pure spinor action
\eqn\pureaction{S = \int d\tau [ P_m \dot x^m + w_\a \dot \l^\a - p_\a \dot \t^\a ].}

Finally, the BRST transformations in this gauge reduce to
\eqn\brstpure{ Q\t^\a = \l^\a, \quad Q x^m =\half \l\g^m \t, \quad Q w_\a =- p_\a -\half P^m (\g_m \t)_\a, \quad Q p_\a = M_\a = \half P^m (\g_m \l)_\a, }
which are the BRST transformations generated by the pure spinor superparticle BRST operator
\eqn\purebrst{ Q = \l^\a d_\a}
where 
\eqn\ddefp{d_\a \equiv p_\a +\half P^m (\g_m\t)_\a}
is the worldsheet version of the
d=10 superspace derivative $D_\a \equiv {{\p}\over{\p\t^\a}} +\half (\g_m\t)_\a {{\p}\over{\p x^m}}$. 
Note that to obtain the pure spinor superparticle action in the usual gauge
\eqn\pureactiontwo{S = \int d\tau [ P_m \dot x^m +\half P_m P^m + w_\a \dot \l^\a - p_\a \dot \t^\a ],}
one needs to add
the BRST-trivial term $- Qb = \half P_m P^m$ to the action of \pureaction\ where 
\eqn\bghost{b = - \half P^m (\bar\xi\g_m d)}
is the pure spinor $b$ ghost and $\bar\xi_\a$ is any spinor satisfying $\l^\a \bar\xi_\a =1$. 
Although it will not be reviewed here, it was explained in \ref\nonmin{N.~Berkovits,
  ``Pure spinor formalism as an N=2 topological string,''
JHEP {\bf 0510}, 089 (2005).
[hep-th/0509120].} how to covariantize the construction of the pure spinor $b$ ghost of \bghost\ by adding a non-minimal sector to the pure spinor formalism.

\subsec{Green-Schwarz superparticle}

In this section, the Green-Schwarz description of the d=10 superparticle will be obtained by
performing a different gauge-fixing of the twistor-like action of \partact. One again gauges
$a=0$ by adding the gauge-fixing fermion $Q (\b a)$ to the classical action.
But instead of gauge-fixing $L^\a=0$ by adding the gauge-fixing term $Q(p_\a L^\a)$, one
chooses the gauge-fixing fermion $\chi$ so that the BRST variation of $w_\a$ vanishes. Since
\eqn\varw{ Q w_\a = - \half P_m (\g^m \t)_\a - \phi {{\d \chi}\over{\d L^\a}},}
this implies that the gauge-fixing fermion $\chi$ is chosen to depend on $L^\a$ as
\eqn\Ldepxi{\chi =\b a -\half \phi^{-1} L^\a P_m (\g^m \t)^\a.} 
As before, the ghost-for-ghost variable $\phi$ will be assumed to be non-vanishing and the scale symmetry of \gaugeA\ will be used to gauge-fix $\phi=1$. Since $\l^\a$ carries scale weight $+1$ and $\t^\a$ carries scale weight $-1$, this shifts the ghost-number
of $\l^\a$ from $0$ to $+1$ and shifts the ghost number of 
$\t^\a$ from $+1$ to $0$.

After this gauge-fixing, the resulting action is
\eqn\gfgu{S= \int d\tau [P_m \dot x^m + w_\a \nabla \l^\a -\half L^\a P_m (\g^m \l)_\a
+ Q(\b a -\half \phi^{-1} L^\a P_m (\g^m \t)_\a)]}
$$= \int d\tau [P_m \dot x^m + w_\a \nabla \l^\a -  L^\a P_m (\g^m \l)_\a + \b\nabla\phi
-\half (\nabla \t^\a - a \l^\a) P_m (\g^m\t)_\a + N a]$$
\eqn\gfgs{= \int d\tau [ P_m \dot x^m + w_\a \dot \l^\a - L^\a P_m (\g^m \l)_\a 
-\half \dot\t^\a P_m (\g^m\t)_\a ]}
where the auxiliary equations of motion $A = N -\half (\l\g_m\t)P^m  = a =\b -\half w_\a\l^\a =0$ have been used.

The BRST transformation in this gauge is
\eqn\brstgs{Q\t^\a = \l^\a, \quad Q x^m =\half \l\g^m \t, \quad \d w_\a =0,\quad \d L^\a = \dot \t^\a,}
and one can easily verify that the Noether charge associated with this transformation vanishes. So in this gauge, the global BRST symmetry is a local gauge symmetry which will be related below to the usual GS kappa symmetry.

To relate the action and symmetries of \gfgs\ and \brstgs\ with the GS action and symmetries,
add to \gfgs\ the BRST-trivial term
\eqn\trivial{ Q [ -w \dot \t + \half (w \g^m \bar\xi)(\l\g_m\dot\t) ] = - w_\a \dot\l^\a}
to remove the $w_\a$ dependence where $\bar\xi_\a$ is a spinor satisfying $\bar\xi_\a \l^\a =1$ and the $\half (w \g^m \bar\xi)(\l\g_m\dot\t)$ term in \trivial\ is needed to preserve
the $w_\a$ gauge invariance of \projw. Then shift
$L^\a \to L^\a - \half P^m (\g_m \bar\xi)$ in \gfgs\
to obtain the action 
\eqn\gstwo{ S = S_{GS} - 
\int d\tau L^\a P_m (\g^m\l)_\a}
where $S_{GS}$ is the standard GS superparticle action in the gauge $e=1$,
\eqn\gsa{S_{GS} = \int d\tau [ P_m (\dot x^m -\half \dot\t\g^m \t) + \half P_m P^m ].}

The role of the second term in \gstwo\ is to impose the
constraint $P^m (\g_m \l)_\a =0$, which implies that $P_m P^m =0$ and that
\eqn\lparam{\l^\a = P^m (\g_m \k)^\a}
 for some $\k_\a$. With this pameterization of $\l^\a$, the BRST transformation of \brstgs\
 becomes the standard GS kappa transformation
 \eqn\gsk{ \d\t^\a = P^m (\g_m\k)^\a, \quad \d x^m =\half \d\t^\a (\g_m\t)_\a}
 and the transformation
 of the second term in \gstwo\ is
 \eqn\transfs{-\d L^\a P_m (\g^m \l)_\a =- (\l\g^m\dot\t)P_m = - (\k\dot\t)P^m P_m.}
So the BRST transformation of $L^\a$ in \brstgs, $Q L^\a = \dot\t^\a$,  replaces the kappa-transformation
of the metric, $\d e = -\k_\a\dot\t^\a$, in the usual GS action.
 
\newsec{ Twistor-like Superstring}

\subsec{Worldsheet action}

In this section, we will generalize the superparticle results of the previous section to the superstring and it will be convenient to work in first-order form with $x^m (\tau,\s)$ and $P_m(\tau,\s)$ variables for the spacetime position and its conjugate momentum. In addition to these spacetime vector variables, the classical 
worldsheet variables for the Type II superstring will contain two sets of projective pure spinors, $\l^\a$ and $\lh^\ah$, and their conjugate momenta, $w_\a$ and $\wh_\ah$, satisfying the pure spinor conditions
\eqn\projpures{ \l \g^m \l =0,  \quad \lh\g^m \lh =0}
and related gauge invariances
\eqn\projws{ w_\a \sim w_\a + \Lambda^m (\g^m \l)_\a, \quad \wh_\ah \sim \wh_\ah + \widehat\Lambda^m (\g^m \lh)_\ah,}
as well as the local scale invariances
 \eqn\scalps{ \l^\a \sim \Omega\l^\a, \quad w_\a \sim \O^{-1} w_\a, \quad \lh^\ah \sim \Oh \lh^\ah, \quad \wh_\ah \sim \Oh^{-1} \wh_\ah }
 where $\O(\tau,\s)$ and $\Oh(\tau,\s)$ are independent local scale parameters and 
 $\a,\ah=1$ to 16 are spinor indices of the same chirality for the Type IIB superstring
 and spinor indices of the opposite chirality for the Type IIA superstring. For the heterotic
 superstring, only one set of projective pure spinor variables is needed and the right-moving sector of the superstring is the same as in the RNS formalism.
 
The twistor-like constraint $P^m (\g_m \l)_\a=0$ for the superparticle has the obvious
generalization for the superstring 
\eqn\tcs{(P^m + \p_\s x^m)(\g_m \l)_\a =0 \quad {\rm and}\quad
(P^m - \p_\s x^m)(\g_m \lh)_\a =0 .}
However, since the commutator
\eqn\comm{[(P^m + \p_\s x^m)(\g_m \l)_\a (\s_1), ~(P^m + \p_\s x^m)(\g_m \l)_\b (\s_2) ] =
\d(\s_1 - \s_2) (\g^m\p_\s\l)_{[\a} (\g_m \l)_{\b]} }
is nonzero, the additional constraints
\eqn\acon{ \nabla_\s \l^\a = 0 \quad {\rm and} \quad \widehat\nabla_\s \lh^\ah =0}
will be imposed where 
\eqn\defnl{\nabla_\s \l^\a \equiv \p_\s \l^\a + A_\s \l^\a, \quad
\widehat\nabla_\s \lh^\ah \equiv \p_\s \lh^\ah + \widehat A_\s \lh^\ah}
and $A_I$ and $\hat A_I$ for $I=\tau,\s$ are worldsheet gauge fields which transform as
$\d A_I = -\O^{-1} \p_I \O$ and $\d \widehat A_I = - \Oh^{-1} \p_I \Oh$ under the local scale
transformations of \scalps.

So the classical worldsheet action for the Type II
superstring is
\eqn\actsupers{S_c = \int d\tau d\s [ P_m \dot x^m + w_\a \nabla_\tau \l^\a +
\wh_\ah \widehat\nabla_\tau \lh^\ah}
$$ -\half L^\a (P_m + \p_\s x_m) (\g^m \l)_\a -\half
\widehat L^\ah (P_m - \p_\s x_m )(\g^m \lh)_\ah + K_\a \N_\s \l^\a + \widehat K_\ah \Nh_\s \lh^\ah]$$
where the Lagrange multipliers $(L^\a, \widehat L^\ah, K_\a, \widehat K_\ah)$ transform under the scale transformations of \scalps\ as
 \eqn\scalps{ L^\a \sim \Omega^{-1}L^\a, \quad K_\a \sim \O^{-1} K_\a, \quad \widehat L^\ah \sim \Oh^{-1} \widehat L^\ah, \quad \widehat K_\ah \sim \Oh^{-1} \widehat K_\ah }
Note that this action is invariant under worldsheet reparameterizations since
the Virasoro constraints 
\eqn\vir{\half(P + \p_\s x)^2 + 2 w_\a \N_\s \l^\a= 0, \quad
\half (P - \p_\s x)^2 -2 \wh_\ah \Nh_\s \lh^\ah =0,}
are implied by the constraints
\eqn\twi{(P_m + \p_\s x_m) (\g^m \l)_\a=0, \quad  \N_\s \l^\a=0, \quad
 (P_m - \p_\s x_m )(\g^m \lh)_\ah =0, \quad \Nh_\s \lh^\ah =0.}
So the change in the kinetic term under worldsheet reparameterizations can be cancelled by an appropriate shift of
the Lagrange multipliers $(L^\a, \widehat L^\ah, K_\a, \widehat K_\ah)$.\foot{Worldsheet reparameterization invariance can be made manifest by shifting the Lagrange multipliers such that the term $e T + \bar e \bar T$ is added to the action of \actsupers\ where $T$ and $\bar T$ are the Virasoro constraints of \vir. Plugging in the auxiliary equation of motion from varying $P_m$,
\eqn\auxp{(e + \bar e) P^m = - \dot x^m + (\bar e- e) \p_\s x^m + \half (L\g^m \l + \widehat L\g^m \lh),}
one obtains
the manifestly worldsheet reparameterization invariant action
\eqn\actsuperscov{S_c = \int dz d\bar z (\det e) [-\half\N_z x^m \N_{\bar z} x_m + 
w_\a \nabla_{\bar z} \l^\a +
\wh_\ah \widehat\nabla_z \lh^\ah}
$$ +\half \N_z x_m (L\g^m \l)_\a +\half
\N_{\bar z} x_m (\widehat L\g^m \lh)  -{1\over 4} (L\g^m \l)(\widehat L\g_m\lh) + 
 K_\a \N_z \l^\a + \widehat K_\ah \Nh_{\bar z} \lh^\ah]$$
where $\nabla_z = e_z^I \nabla_I$ and $\nabla_{\bar z} = e_{\bar z}^I \nabla_I$ for
$I = (\tau, \sigma)$, and $(e_z^I, e_{\bar z}^I)$ is the worldsheet vierbein which is
defined (up to an SO(1,1) rotation and conformal transformation) as
\eqn\vier{e_z^\tau = 1, \quad e_z^\sigma =-2 \bar e, 
\quad e_{\bar z}^\tau = 1, \quad e_{\bar z}^\sigma = 2 e.}}

In addition to the worldsheet reparameterizations and local scale symmetries, the action is invariant under the local symmetries related to the first-class constraints of \twi,
\eqn\ssym{\d x^m =\half( \l\g^m\t + \lh\g^m\th), \quad \d P_m = \half\p_\s (\l\g^m\t - \lh\g^m\th), }
$$\d w_\a =-\half (P_m + \p_\s x_m )(\g^m\t)_\a - \N_\s c_\a - {{\d \chi}\over{\d L^\a}} \phi -
{1\over 4} {{\d \chi}\over{\d K^\b}} (\g^m\t)_\b
(\g_m\t)_\a, $$
$$\d \wh_\ah =-\half (P_m - \p_\s x_m )(\g^m\th)_\ah - \Nh_\s \widehat c_\ah - {{\d \chi}\over{\d \widehat L^\ah}}\widehat \phi +{1\over 4}  {{\d \chi}\over{\d \widehat K^\bh}} (\g^m\th)_\bh
(\g_m\th)_\ah,$$
$$\d L^\a = \N_\tau\t^\a - a \l^\a, \quad\d K_\a = \N_\tau c_\a +{1\over 4} (L\g^m\l)(\g_m\t)_\a - {1\over 4}(\t\g^m\l)(\g_m L)_\a, $$
$$ \d \widehat L^\ah = \Nh_\tau \th^\ah - \widehat a \lh^\ah, \quad
\d \widehat K_\ah = \Nh_\tau \widehat c_\ah -{1\over 4} (\widehat L\g^m\lh)(\g_m\th)_\ah +{1\over 4} (\th\g^m\lh)(\g_m \widehat L)_\ah,$$
$$\d c_{\a\b} = -{1\over 4}
(\t\g^m)_\b(\g_m\t)_\a, \quad \d \t^\a = \phi \l^\a, \quad \d a= \N_\tau \phi,$$
$$ \d\widehat c_{\ah\bh} = {1\over 4} (\th\g^m)_\bh(\g_m\th)_\ah,
\quad \d \th^\ah = \widehat\phi \lh^\ah, \quad \d\widehat a =  \Nh_\tau\widehat\phi,$$
where
\eqn\defcch{c_\a \equiv c_{\a\b}\l^\b, \quad \widehat c_\ah \equiv \widehat c_{\ah\bh}\lh^\bh, \quad c_{\a\b} = -c_{\b\a}, \quad \widehat c_{\ah\bh} = -\widehat c_{\bh\ah}}
so that $c_\a \l^\a =0$ and $\widehat c_\ah \lh^\ah =0$ \foot{This restriction on $c_\a$ and $\widehat c_\ah$ is necessary since $\d S_c =
\int d\tau d\s (\l^\a c_\a + \lh^\ah \widehat c_\ah) (\p_\tau A_\s-\p_\s A_\tau)$. I would like
to thank Sebastian Guttenberg for pointing out this mistake in the original version of the paper.},
$\chi$ is the gauge-fixing fermion, and the variations of the ghosts $(c_{\a\b}, \widehat c_{\ah\bh}, \t^\a, \th^\ah, a, \widehat a)$ in the last
two lines of \ssym\ come from the nonvanishing commutator of \comm\ and from the gauge-for-gauge symmetries related to \gfg. As in \tranchi, the shifts in $w_\a$ and
$\widehat w_\ah$ proportional to the gauge-fixing fermion $\chi$ are needed to cancel
terms proportional to the equations of motion $\N_\tau \l^\a$ and $\Nh_\tau \lh^\ah$ in the gauge-for-gauge transformations of $(\d K_\a, \d L^\a)$ and $(\d\widehat K_\ah, \d\widehat L^\ah)$.

\subsec{Pure spinor superstring}

As in the superparticle, gauge-fixing to the pure spinor formalism is achieved by first
using the local symmetries of \ssym\ to gauge $L^\a = \widehat L^\ah =0$ and $a = \widehat a =0$.
Using the BRST method, this involves adding to the classical action $S_c$ of \actsupers\ the gauge-fixing fermion 
\eqn\gff{\chi = p_\a L^\a + \widehat p_\ah \widehat L^\ah + \b a 
+\widehat\b \widehat a }
$$- \phi^{-1} (K_\a \N_\s\t^\a +L^\a \N_\s c_\a) - \widehat\phi^{-1}( \widehat K_\ah \Nh_\s \th^\ah    
 +\widehat L^\ah \Nh_\s \widehat c_\ah ) $$
 $$-\half (\p_\tau A_\s -\p_\s A_\tau) (\phi^{-1} c_{\a\b}\t^\a\t^\b + \widehat\phi^{-1}
 \widehat c_{\ah\bh}\th^\ah\th^\bh),$$
where the second and third lines of \gff\ have been included to eliminate $(K_\a, \widehat K_\ah)$ and $(c_\a, \widehat c_\ah)$ from the action. The resulting gauge-fixed action is
\eqn\gfss{S = S_c + }
$$\int d\tau d\s Q[p_\a L^\a + \widehat p_\ah \widehat L^\ah + \b a +
\widehat\b \widehat a - \phi^{-1}(K_\a \N_\s\t^\a +L^\a \N_\s c_\a)-\widehat\phi^{-1}( \widehat K_\ah \Nh_\s \th^\ah +  
\widehat L^\ah \Nh_\s \widehat c_\ah )$$
$$-\half (\p_\tau A_\s -\p_\s A_\tau) (\phi^{-1} c_{\a\b}\t^\a\t^\b + \widehat\phi^{-1}
 \widehat c_{\ah\bh}\th^\ah\th^\bh)]$$
$$ = S_c + \int d\tau d \s [ -p_\a \N_\tau \t^\a  + M_\a L^\a +  \b \N_\tau \phi + (N + \l^\a p_\a - \phi^{-1}\l^\a\N_\s c_\a) a $$
$$- \widehat p_\ah \Nh_\tau \th^\ah  +\widehat M_\ah \widehat L^\ah +\widehat\b
\Nh_\tau \widehat\phi + (\widehat N +\lh^\ah \widehat p_\ah - \widehat\phi^{-1}\lh^\ah \Nh_\s \widehat c_\ah) \widehat a$$
$$- K_\a \N_\s \l^\a -{1\over 4} \phi^{-1} (L\g^m\t) (\t\g_m\N_\s\l)  -{3\over 4}
\phi^{-1}(L\g^m\N_\s\t) (\l\g_m\t)) $$
$$- \widehat K_\ah \Nh_\s \lh^\ah +{1\over 4}\widehat\phi^{-1} (\widehat L^\ah\g^m\th)(\th\g_m\Nh_\s\lh)  +{3\over 4} \widehat\phi^{-1}
(\widehat L\g^m\Nh_\s\th)(\lh\g_m\th)]$$
where $(M_\a, \widehat M_\ah)$ and
$(N, \widehat N)$ are the Nakanishi-Lautrup fields defined by 
\eqn\nlf{Q p_\a = M_\a, \quad Q \widehat p_\a = \widehat M_\ah, \quad Q \b = N, \quad 
Q\widehat \b = \widehat N.}

The next step is to use the local scale symmetries to gauge-fix $\phi = \widehat\phi =1$, where it
will be assumed that these fields are non-vanishing on the entire worldsheet. As in
the superparticle, this shifts the ghost number by the scale weight so that after gauge-fixing
$\phi = \widehat\phi =1$, the pure spinor variables $(\l^\a, w_\a)$
and $(\lh_\ah, \widehat w_\ah)$ carry ghost number $(1,-1)$ and the fermionic spacetime
spinor variables
$(\t^\a, p_\a)$ and
$(\th^\ah, \widehat p_\ah)$ carry ghost number $(0,0)$.

After solving the auxiliary equations of motion for the variables 
$(A_\tau, \b, M_\a, L^\a, N, a)$ and
$(\widehat A_\tau, \widehat\b, \widehat M_\ah, \widehat L^\ah, \widehat N, \widehat a)$, the action reduces to
\eqn\reducess{S = \int d\tau d\s [ P_m \dot x^m + w_\a \dot \l^\a + \wh_\ah \dot\lh^\ah
 - p_\a \dot \t^\a - \widehat p_\ah \dot \th^\ah],}
with the BRST transformations
\eqn\brstss{Q \t^\a = \l^\a, \quad Q \th^\ah = \lh^\ah, \quad Q x^m =\half( \l\g^m\t + \lh \g^m\th),\quad  Q P_m =\half \p_\s (\l\g^m\t - \lh\g^m\th), } 
$$Q w_\a =-\half (P_m + \p_\s x_m )(\g^m\t)_\a - p_\a +{1\over 4} (\g_m\t)_\a (\p_\s\t \g^m \t), $$
$$
Q p_\a =\half (P_m + \p_\s x_m )(\g^m\l)_\a +{3\over 4} (\g_m\p_\s\t)_\a (\l\g^m \t) +
{1\over 4} (\g_m\t)_\a (\t\g^m \p_\s \l), $$
$$Q \wh_\ah =-\half (P_m - \p_\s x_m )(\g^m\th)_\ah -\widehat p_\ah -{1\over 4} (\g_m\th)_\ah (\p_\s\th \g^m \th), $$
$$
Q \widehat p_\ah =\half (P_m - \p_\s x_m )(\g^m\lh)_\ah -{3\over 4} (\g_m\p_\s\th)_\ah (\lh\g^m  \th) -{1\over 4} (\g_m\th)_\ah (\th\g^m\p_\s \lh) .$$
These are the usual pure spinor BRST transformations generated by the BRST operator
\eqn\usualb{Q = \int d\s (\l^\a d_\a + \lh^\ah \widehat d_\ah)}
where 
\eqn\deriv{d_\a \equiv p_\a +\half (P_m + \p_\s x_m )(\g^m\t)_\a -{1\over 4} (\g_m\t)_\a (\p_\s\t \g^m \t),}
$$\widehat d_\ah \equiv \widehat p_\ah +\half (P_m - \p_\s x_m )(\g^m\th)_\ah +{1\over 4} (\g_m\th)_\ah (\p_\s\th \g^m \th)$$
are the worldsheet versions of the N=2 d=10
superspace derivatives $D_\a \equiv {{\p}\over{\p\t^\a}} +\half (\g_m\t)_\a {{\p}\over{\p x^m}}$ and
$\widehat D_\ah = {{\p}\over{\p\th^\ah}} +\half (\g_m\th)_\ah {{\p}\over{\p x^m}}$.

To relate \reducess\ with the usual pure spinor superstring action in conformal gauge, one needs
to add the BRST-trival term $-\half \int d\tau d\s Q (b + \bar b)$ where 
\eqn\bgs{ b =-\half (\bar\xi \g^m d) (P_m + \p_\s x_m + \t\g_m\p_\s \t) -2 w\p_\s\t +
(w\g_m\bar\xi)(\l\g^m\p_\s\t), }
$$\widehat b = -\half  (\bar{\widehat\xi} \g^m \widehat d) (P_m - \p_\s x_m-\th\g_m\p_\s\th) +
2 \wh \p_\s \th^\ah  -(\wh\g_m\bar{\widehat\xi})(\lh\g^m \p_\s \th)$$
are the pure spinor $b$ and $\widehat b$ ghosts, and $\bar\xi_\a$ and $\widehat{\bar\xi}_\ah$ are
spinors satisfying $\bar\xi_\a \l^\a =1$ and $\widehat{\bar\xi}_\ah \lh^\ah=1$.
One can verify that $Qb = T$ and $Q\widehat b = \widehat T$ where 
\eqn\stress{ T= -\half (P^m + \p_\s x^m)(P_m + \p_\s x_m) -2 w_\a \p_\s \l^\a +2 p_\a \p_\s\t^\a,}
$$\widehat T= -\half (P^m - \p_\s x^m)(P_m - \p_\s x_m) +2 \wh_\a \p_\s \lh^\ah -2 \widehat p_\ah \p_\s\th^\a
$$
are the left and right-moving stress-tensors.
After adding $-\half\int d\tau d\s Q (b + \bar b)$  to \reducess\ and integrating out $P_m$, one obtains the conformally invariant pure spinor worldsheet action 
\eqn\confp{S = \int d\tau d\s [-\half \p x_m \bar\p x^m + w_\a \bar\p \l^\a + \wh_\ah \p\lh^\ah
 - p_\a \bar\p \t^\a - \widehat p_\ah \p \th^\ah],}
 where $\p = \p_\tau - \p_\s$ and $\bar\p = \p_\tau + \p_\s$.

\subsec{Green-Schwarz superstring}

As in the superparticle, gauge-fixing to the GS formalism for the superstring is achieved by
gauging $a=\widehat a=0$ and choosing the $L$ dependence of the gauge-fixing fermion $\chi$ such that the BRST
variation of $w_\a$ and $\widehat w_\ah$ vanishes where
\eqn\brstw{Q w_\a = -\half(P_m + \p_\s x_m )(\g^m\t)_\a - \N_\s c_\a - {{\d \chi}\over{\d L^\a}} \phi -{1\over 4} {{\d \chi}\over{\d K^\b}} (\g^m\t)_\b
(\g_m\t)_\a, }
$$Q \wh_\ah = -\half(P_m - \p_\s x_m )(\g^m\th)_\ah - \Nh_\s \widehat c_\ah - {{\d \chi}\over{\d \widehat L^\ah}}\widehat \phi +{1\over 4}  {{\d \chi}\over{\d \widehat K^\bh}} (\g^m\th)_\bh
(\g_m\th)_\ah.$$

To gauge-fix $a=\widehat a=0$ and remove the $(K_\a, \widehat K_\ah)$ dependence from the
action, the $L$-independent terms in $\chi$ will be chosen to be
\eqn\nol{\b a + \widehat\b \widehat a - \phi^{-1} K_\a \N_\s \t^\a - \phi^{-1}\widehat K_\ah \N_\s\th^\ah.}
The vanishing of $Q w_\a$ and $Q \wh_\ah$ therefore implies that 
the gauge-fixing fermion is
\eqn\gfgs{\chi = \b a + \widehat\b \widehat a - \phi^{-1} K_\a \N_\s \t^\a - \phi^{-1}\widehat K_\ah \N_\s\th^\ah}
$$+ \phi^{-1} L^\a [-\half (P^m + \p_\s x^m) (\g_m\t)_\a - \N_\s c_\a +{1\over 4}\phi^{-1} (\N_\s\t\g^m \t)(\g_m\t)_\a ]$$
$$ + \widehat\phi^{-1} \widehat L^\ah [-\half (P^m - \p_\s x^m) (\g_m\th)_\ah - \Nh_\s \widehat c_\ah -{1\over 4}\widehat\phi^{-1} (\Nh_\s\th\g^m \th)(\g_m\th)_\ah ],$$
and the resulting gauge-fixed
action is 
\eqn\gfss{S = S_c + 
\int d\tau d\s Q[\b a + \widehat\b \widehat a - \phi^{-1} K_\a \N_\s \t^\a - \phi^{-1}\widehat K_\ah \N_\s\th^\ah}
$$+ \phi^{-1} L^\a [-\half (P^m + \p_\s x^m) (\g_m\t)_\a - \N_\s c_\a +{1\over 4}\phi^{-1} (\N_\s\t\g^m \t)(\g_m\t)_\a ]$$
$$ + \widehat\phi^{-1} \widehat L^\ah [-\half (P^m - \p_\s x^m) (\g_m\th)_\ah - \Nh_\s \widehat c_\ah -{1\over 4}\widehat\phi^{-1} (\Nh_\s\th\g^m \th)(\g_m\th)_\ah ]]$$
$$ = S_c+\int d\tau d \s [
- K_\a \N_\s \l^\a -\half \phi^{-1}(P^m + \p_\s x^m) (\N_\tau\t\g_m\t)  +{1\over 4}\phi^{-2} (\N_\s\t\g^m \t)(\N_\tau\t\g_m\t)$$
$$-\widehat K_\ah \Nh_\s \lh^\ah -\half\widehat\phi^{-1}  (P^m - \p_\s x^m) (\Nh_\tau\th\g_m\th)  -{1\over 4}\widehat\phi^{-2} (\Nh_\s\th\g^m \th)(\Nh_\tau\th\g_m\th)$$
$$-\half  
(L\g^m\l) (P_m + \p_\s x_m + 2\phi^{-1}\t\g_m\p_\s\t) -\half(\widehat L\g^m\lh) (P_m - \p_\s x_m -2 \widehat\phi^{-1}\th\g_m\p_\s\th)$$
$$
 +\b \N_\tau \phi + (N - \phi^{-1}\l^\a\N_\s c_\a
-\half\phi^{-1}  (P^m + \p_\s x^m) (\l\g_m\t)_\a  +{1\over 4}\phi^{-2} (\N_\s\t\g^m \t)(\l\g_m\t)) a $$
$$+ \widehat\b
\Nh_\tau \widehat\phi + (\widehat N  - \widehat\phi^{-1}\lh^\ah \Nh_\s \widehat c_\ah
-\half\widehat\phi^{-1}  (P^m - \p_\s x^m) (\lh\g_m\th)_\ah  -{1\over 4}\widehat\phi^{-2} (\Nh_\s\th\g^m \th)(\lh\g_m\th)) \widehat a]$$
$$ = \int d\tau d\s [ P^m \dot x_m + w_\a \N_\tau \l^\a + \wh_\ah\Nh_\tau\lh^\ah 
+\b \N_\tau \phi + \widehat\b
\Nh_\tau \widehat\phi $$
$$-\half \phi^{-1} (P^m + \p_\s x^m)
(\N_\tau\t \g_m \t) -\half \widehat\phi^{-1}(P^m - \p_\s x^m)(\Nh_\tau\th \g_m\th)$$
$$+{1\over 4}\phi^{-2}(\N_\s\t\g^m\t)(\N_\tau \t \g_m \t) 
-{1\over 4}\widehat\phi^{-2}(\Nh_\s\th\g^m\th)(\Nh_\tau \th \g_m \th) $$
$$ -(L\g^m\l) (P_m + \p_\s x_m + \phi^{-1}\t\g_m\p_\s\t) -(\widehat L\g^m\lh) (P_m - \p_\s x_m - \widehat\phi^{-1}\th\g_m\p_\s\th)$$
$$ + (N - \phi^{-1}\l^\a\N_\s c_\a
-\half\phi^{-1}  (P^m + \p_\s x^m) (\l\g_m\t)_\a  +{1\over 4}\phi^{-2} (\N_\s\t\g^m \t)(\l\g_m\t)) a $$
$$+ (\widehat N  - \widehat\phi^{-1}\lh^\ah \Nh_\s \widehat c_\ah
-\half\widehat\phi^{-1}  (P^m - \p_\s x^m) (\lh\g_m\th)_\ah  -{1\over 4}\widehat\phi^{-2} (\Nh_\s\th\g^m \th)(\lh\g_m\th)) \widehat a ]$$
where $N = Q\b$ and $\widehat N = Q\widehat\b$ are Nakanishi-Lautrup fields.

After using the scale symmetries to gauge-fix $\phi = \widehat \phi =1$ and applying the auxiliary equations of motion for $(A_\tau, \b, a, N)$ and $(\widehat A_\tau, \widehat\b, \widehat a, \widehat N)$, the action of \gfss\ simplifies to 
\eqn\simpgs{ S= \int d\tau d\s [ P^m \dot x_m + w_\a \dot \l^\a + \wh_\ah\dot\lh^\ah}
$$ 
-\half (P^m + \p_\s x^m -\half \p_\s\t\g^m \t)
(\dot\t \g_m \t) -\half (P^m - \p_\s x^m +\half \p_\s \th\g^m\th)(\dot\th \g_m\th)$$
$$ -(L\g^m\l) (P_m + \p_\s x_m + \t\g_m\p_\s\t) -(\widehat L\g^m\lh) (P_m - \p_\s x_m -\th\g_m\p_\s\th)].$$
As in the superparticle action, one can relate \simpgs\ to the GS superstring action in conformal gauge
by adding the BRST-trivial term
\eqn\trivials{Q [ - w\dot\t +\half (w\g^m \bar\xi)(\l\g_m\dot\t) - \wh\dot\th +
\half (\wh\g^m \widehat{\bar\xi})(\lh\g_m\dot\th)] = - w_\a \dot\l^\a
 -\wh_\ah\dot\lh^\ah}
and shifting
\eqn\lshift{L^\a \to L^\a -{1\over 4} (P^m + \p_\s x^m +\t\g^m\p_\s\t) (\g_m \bar\xi)_\a, \quad
\widehat L^\ah \to \widehat L^\ah -{1\over 4} (P^m - \p_\s x^m -\th\g^m\p_\s\th) (\g_m \widehat{\bar\xi})_\ah}
where $\bar\xi_\a$ and $\widehat{\bar\xi}_\ah$ are spinors satisfying $\bar\xi_\a \l^\a =1$
and $\widehat{\bar\xi}_\ah \lh^\ah =1$.  One obtains  
\eqn\gsthree{  
S= \int d\tau d\s [ P^m \dot x_m -\half  (P^m + \p_\s x^m-\half \p_\s\t^m \t)(\dot\t \g_m\t)
-\half (P^m - \p_\s x^m +\half \p_\s\th\g^m \th)(\dot\th \g_m\th)}
$$+ {1\over 4} (P^m + \p_\s x^m +\t\g^m\p_\s\t) (P_m + \p_\s x_m +\t\g_m\p_\s\t)$$
$$+{1\over 4}
 (P^m - \p_\s x^m -\th\g^m\p_\s\th) (P^m - \p_\s x^m -\th\g^m\p_\s\th)$$
$$ -(L\g^m\l) (P_m + \p_\s x_m + \t\g_m\p_\s\t) -(\widehat L\g^m\lh) (P_m - \p_\s x_m -\th\g_m\p_\s\th)],$$
which after integrating out $P_m$ is equal to
\eqn\gsfour{S= S_{GS} + \int d\tau d\s [ 
(L\g^m\l) \Pi_m +(\widehat L\g^m\lh) \bar\Pi_m - (L\g^m\l)(\widehat L\g_m \lh) ]}
where 
$S_{GS}$ is the GS action in conformal gauge
\eqn\confgs{S_{GS} = \int d\tau d\s [ -\half\Pi^m \bar\Pi_m -\half \p_\tau x_m (\t \g^m \p_\s\t - \th\g^m \p_\s\th) +\half \p_\s x_m
(\t\g^m \p_\tau\t - \th\g^m \p_\tau\th) }
$$ +{1\over 4} (\t\g^m\p_\tau\t)(\th\g_m\p_\s\th) - 
{1\over 4} (\t\g^m\p_\s\t)(\th\g_m\p_\tau\th)],$$
$\p\equiv \p_\tau - \p_\s$ and $\bar\p \equiv \p_\tau + \p_\s$, and
$$\Pi^m = \p x^m + \half (\t\g^m\p\t +\th\g^m \p\th), \quad \bar\Pi^m = \bar\p x^m +\half( \t\g^m \bar\p\t +\th \g^m \bar\p\th). $$

As in the superparticle, the role of the second term in \gsfour\ is to impose the constraints
$\Pi^m (\g_m\l)_\a=0$ and $\bar\Pi^m (\g_m\lh)_\ah=0$, which implies that 
$\Pi_m\Pi^m = \bar\Pi_m \bar\Pi^m =0$ and that
\eqn\lgsp{\l^\a = \Pi^m (\g_m\k)^\a, \quad \lh^\ah = \bar\Pi^m (\g_m \widehat\k)^\ah}
for some $\k_\a$ and $\widehat\k_\ah$. With this parameterization of $\l^\a$ and $\lh^\ah$,
the BRST transformations of \ssym\ reduce to the usual GS kappa transformations
 \eqn\gssk{ \d\t^\a = \Pi^m (\g_m\k)^\a, \quad
 \d\th^\ah = \bar\Pi^m (\g_m\kh)^\ah, \quad \d x^m =\half( \d\t\g^m\t + \d\th\g^m\th),}
 $$\d \Pi^m = \d\t\g^m \p\t + \d\th\g^m\p\t, \quad \d\bar\Pi^m = \d\t\g^m\bar\p\t + \d\th\g^m\bar\p\th.$$
 Using the BRST transformation of the shifted $L^\a$ and $\widehat L^\ah$ in \lshift, 
 \eqn\tranL{\d L^\a = \dot \t^\a +\half (\l\g^m\p_\s\t)(\g_m\bar\xi)^\a, \quad
 \d \widehat L^\ah = \dot \th^\ah -\half (\lh\g^m\p_\s\th)(\g_m\widehat{\bar\xi})^\ah,}
 the kappa transformation of the $L$-dependent term in \gsfour\ is
 \eqn\transfs{\int d\tau d\s [ \Pi_m (\l \g^m \bar\p\t)_\a + \bar\Pi_m (\lh\g^m \p\t)] =
 \int d\tau d\s [  \Pi^m \Pi_m (\k\bar\p\t) + \bar\Pi^m \bar\Pi_m (\widehat\k \p\th)],}
which cancels the kappa transformation of the GS action in conformal gauge.
So the BRST transformations of $L^\a$ and $\widehat L^\ah$ in \tranL\ replace the kappa transformation of the two-dimensional vierbein, $\d e = \k_\a\bar\p\t^\a$ and
$\d \widehat e = \widehat\k_\ah \p\th^\ah$, in the usual reparameterization-invariant GS action.
 
\newsec{Twistor-like Supermembrane}

In this final section, a purely bosonic classical action for the d=11 supermembrane will
be proposed which reduces under double-dimensional reduction to the purely bosonic
classical action for the d=10 Type IIA superstring. Although it will not be attempted here, it should be possible to follow the
same procedure as in the previous sections to gauge-fix
this classical twistor-like action to the pure spinor \bmem\ and GS \town\ versions of the d=11 supermembrane action.

\subsec{ d=11 superparticle}

Just as the GS or pure spinor superparticle in ten dimensions describes d=10 super-Maxwell,
the GS or pure spinor superparticle in eleven dimensions describes d=11 supergravity. 
This was verified in light-cone gauge for the d=11 GS superparticle, and was verified covariantly for the d=11 pure
spinor superparticle in \bmem\memem\cmem.
The classical twistor-like action for the d=11 superparticle is the obvious generalization of \partact\ and is given by
\eqn\elev{S_c = \int d\tau [P_M \dot x^M + w_B \nabla \l^B - \half L^B P_M (\g^M \l)_B]}
where $M=0$ to 10 is an SO(10,1) vector index, $B=1$ to 32 is an SO(10,1) spinor index,
$\l^B$ is a d=11 projective `semi-pure' spinor satisfying the identity and local scale symmetry
\eqn\semipure{\l^B (\tau)\g^M_{BC} \l^C (\tau) = 0, \quad \l^B(\tau) \sim \O(\tau)\l^B(\tau),}
$w_B$ is its conjugate momentum satisfying the gauge symmetry and scale symmetry
\eqn\welev{\d w_B = \L^M (\g_M\l)_B, \quad w_B \sim \O^{-1} w_B,}
$\nabla \l^B = \dot\l^B + A \l^B$ where $A$ is a worldline gauge field transforming as
$A \to A - \O^{-1} \dot \O$ under the local scale
transformation, and $L^B$ is the Lagrange multiplier for the twistor-like constraint
$P_M (\g^M\l)_B =0$. Note that a d=11 projective `semi-pure' spinor satisfies $\l\g^M\l=0$ and has 22 independent
components, whereas a d=11 projective pure spinor would satisfy both $\l\g^M\l=0$ and $\l\g^{MN}\l=0$ and contain 15 independent components. 

Following the same procedure as in section 2, the action of \elev\ can be gauge-fixed
either to the d=11 pure spinor superparticle action
\eqn\elevsp{S = \int d\tau [\dot x^M P_M + w_B \dot \l^B - p_B \dot \t^B]}
with the BRST charge 
\eqn\elevbrst{Q = \l^B (p_B +\half P^M (\g_M\t)_B),}
or to the d=11
GS superparticle action
\eqn\elevgsp{S = \int d\tau [ P_M (\dot x^M +\half \t\g^M\dot\t) +\half P^M P_M - L^B P_M (\g^M\l_B)]}
with the kappa transformation
\eqn\kappaelev{\d\t^B = P^M (\g_M\k)^B, \quad \d x^M =\half \d\t\g^M \t, \quad \d L^M = \dot\t^B,}
where $\l^B = P^M (\g_M\k)^B$.

\subsec{d=11 supermembrane}

The purely bosonic classical  d=11 supermembrane action will involve the worldvolume variables
$(x^M, P_M)$ and $(\l^B, w_B)$ together with the twistor-like constraint
\eqn\elevt{P_M (\g^M\l)_B + \half\e^{jk} \p_j x_M \p_k x_N (\g^{MN}\l)_B =0}
where $\p_j = \p_{\s^j}$ for $j=1,2$ and $(\tau,\s_1,\s_2)$ are the coordinates of the
worldvolume.  Using the d=11 $\g$-matrix identity $(\g_M)_{(BC} (\g^{MN})_{DE)} =0$, one
finds that 
the commutator of the constraint of \elevt\ with itself closes to an algebra if one also imposes
the additional constraints
\eqn\addeleven{\N_j\l^B =0\quad {\rm and} \quad (\l\g^{MN}\l) \p_j x_M =0}
for $j=1,2$ where $\N_j\l^B \equiv \p_j \l^B + A_j \l^B$ and $(A_\tau, A_1, A_2)$ is a
worldvolume gauge field for the scale symmetry.

The twistor-like version of the d=11 supermembrane action will therefore be defined as
 \eqn\twelev{S = \int d\tau d\s^1 d\s^2 [P_M \dot x^M + w_B \nabla_\tau \l^B }
 $$ -\half L^B (
 P_M (\g^M \l)_B + \half\e^{jk} \p_j x_M \p_k x_N (\g^{MN}\l)_B ) + K_B^j \N_j \l^B +
 J_M^j (\l\g^{MN}\l) \p_j x^N ],$$
 where $L^B$, $K^j_B$ and $J_M^j$ are Lagrange multipliers for the constraints of
 \elevt\ and \addeleven. Although it should be possible to verify \twelev\ by gauge-fixing
 it to the pure spinor and GS supermembrane actions of \bmem\ and \town, it will
 instead be verified here by performing a double-dimensional reduction and comparing
 with the twistor-like Type IIA superstring action of the previous section.
 
 Under double-dimensional reduction, the $x^{10}$ coordinate of d=11 spacetime
 is compactified on a circle of circumference $2\pi$, and the $\s_2$ direction of
 the worldvolume is wrapped around this circle. One can choose the parameterization
 $\s_2 = x^{10}$, which implies that
 \eqn\param{\p_\tau x^{10} = \p_1 x^{10} =P^{10} = 0, \quad \p_2 x^{10} =1, \quad
 \p_2 x^m = \p_2 P^m = \N_2 \l^B = \N_2 w_B =0}
 for $m=0$ to 9 and $B=1$ to 32. With this parameterization of $\s_2$, the constraints
 of \elevt\ and \addeleven\ reduce to
 \eqn\redel{P_m (\g^m \l)_B + \p_1 x_m (\g^m\g^{10}\l)_B =0,}
 $$ \N_1\l^B =0,  \quad (\l\g^{mn}\l) \p_1 x^n =0, \quad (\l\g^m \g^{10}\l) =0.$$
 
 Splitting $\l^B \to (\l^\a, \lh_\a)$ for $\a=1$ to 16 where
 $\l^\a = \l^B + (\g^{10}\l)^B$ and $\lh_\a = \l_B - (\g^{10}\l)_B$ are d=10
 Weyl and anti-Weyl spinors, the constraints of \redel\ take the form
 \eqn\formel{ (P_m + \p_1 x_m)(\g^m\l)_\a = 0, \quad (P_m - \p_1 x_m) (\g^m\lh)^\a =0,}
 $$\N_1 \l^\a = \N_1 \lh_\a =0,  \quad (\l\g_{mn}\lh)\p_1 x^n=0, \quad \l\g^m\l - \lh\g^m\lh =0,$$
 Furthermore, the d=11 semi-pure spinor constraint $\l^A \g^M_{AB} \l^B =0$ implies
 that $\l^\a$ and $\lh_\ah$ satisfy
 \eqn\semip{\l\g^m\l + \lh\g^m\lh=0 \quad {\rm and} \quad \l^\a \lh_\a =0.}   
 Since $\l^\a\lh_\a=0$ together with the first line of \formel\
implies that $(\l\g_{mn}\lh)\p_1 x^n=0$, the above constraints imply 
\eqn\tenele{\l\g^m\l=0, \quad \lh\g^m\lh=0, \quad \N_1\l^\a = \N_1 \lh_\a =0,}
$$  (P_m + \p_1 x_m)(\g^m\l)_\a = 0, \quad (P_m - \p_1 x_m) (\g^m\lh)^\a =0,$$
which are precisely the twistor-like constraints of \twi\ for the d=10 Type IIA superstring.

The only subtlety is that the d=11 semi-pure spinor constraint implies that the zero modes of the left and right-moving Type IIA pure spinors, $\l_{(0)}^\a$ and $\lh_{(0)\a}$, satisfy the constraint $\l_{(0)}^\a \lh_{(0)\a}=0$
which is not a constraint of the d=10 Type IIA superstring. However, it will now be argued that
because $\l_{(0)}^\a\lh_{(0)\a}$ is BRST-trivial, the only effect of this new constraint is to trivially double the cohomology of the Type IIA
superstring. To see this, note that the new constraint modifies the Type IIA superstring BRST operator to
\eqn\brsttwo{Q' = \int d\s  (\l^\a d_\a + \lh_\a \widehat d^\a) +  r \l_{(0)}^\a \lh_{(0)\a}}
where $r$ is a constant fermionic ghost for the constraint and
$Q =\int d\s  (\l^\a d_\a + \lh_\a \widehat d^\a)$ is the original Type IIA BRST operator of \usualb.
Since $ r \l^\a\lh_\a = [ Q ,  r ~\t^\a \lh_\a ]$, one can write 
\eqn\simtwo{ Q' = e^{-r~ \t_{(0)}^\a \lh_{(0)\a}} Q  e^{+r ~\t_{(0)}^\a \lh_{(0)\a}}.} 
Defining $r$ to annihilate the ground state and its conjugate momentum $s$ to 
satisfy $\{ r, s\} =1$, \simtwo\ implies that
any state $V'$ in the cohomology of $Q'$ satisfies
\eqn\vnew{  e^{+r~ \t_{(0)}^\a \lh_{(0)\a}} V' e^{-r~ \t_{(0)}^\a \lh_{(0)\a}} = V_1 + s V_2  }
where $V_1$ and $V_2$ are states in the cohomology of $Q$. So the cohomology of $Q'$ is 
a trivial doubling of the Type IIA superstring cohomology because of the extra fermionic ghost zero
mode of $s$. If one wants to obtain a single copy of the Type IIA superstring cohomology, one can restrict to states $V'$ satisfying $[ r, V' ] =0$, which is analogous to the usual $[ (b_{(0)}-\bar b_{(0)}), V ]$ =0 condition in bosonic closed string theory.

\vskip 15pt
{\bf Acknowledgements:}
I would like to thank 
Michael Green, Andrei Mikhailov, Nikita Nekrasov, George Thompson, Cumrun Vafa, Pedro Vieira and Edward Witten for useful discussions, and
CNPq grant 300256/94-9 and FAPESP grants 2009/50639-2 and 2011/11973-4
for partial financial support.

\listrefs

\end